\documentclass[aps,prb,twocolumn,showpacs]{revtex4}
\usepackage{txfonts}
\usepackage{graphicx}

\begin{document}

\title{Physical properties of noncentrosymmetric superconductor Ru$_7$B$_3$}

\author{Lei Fang, Huan Yang, Xiyu Zhu, Gang Mu, Zhao-Sheng Wang, Lei Shan, Cong Ren and Hai-Hu
Wen}\email{hhwen@aphy.iphy.ac.cn} \affiliation{National Laboratory
for Superconductivity, Institute of Physics and National Laboratory
for Condensed Matter Physics, Chinese Academy of Sciences, P. O.~Box
603, Beijing 100190, P.~R.~China}

\date{\today}

\begin{abstract}
Transition metal boride  Ru$_7$B$_3$ was found to be a
noncentrosymmetric superconductor with $T_{C}$ equal to 3.3 K.
Superconducting and normal state properties of Ru$_7$B$_3$ were
determined by a self-consistent analysis through
resistivity($\rho_{xx}$ and $\rho_{xy}$) , specific heat, lower
critical field measurement and electronic  band structure
calculation. It is found that Ru$_7$B$_3$ belongs to an s-wave
dominated single band superconductor with energy gap 0.5 meV and
could be categorized into type II superconductor with weak
electron-phonon coupling. Unusual 'kink' feature is clearly observed
in field-broadening resistivity curves, suggesting the possible
mixture of spin triplet induced by the lattice without inversion
symmetry.
\end{abstract}

\maketitle

\section{Introduction}

Unconventional superconductors have been extensively studied during
the past decades for the underlying fundamental physics or the
potential industrial applications. Some well known examples are
heavy fermion superconductors\cite{heavyfermion},
cuprates\cite{cuprate}, Sr$_{2}$RuO$_{4}$\cite{SrRuO} and the newly
discovered iron arsenide\cite{iron}. In recent years superconductors
lacking of lattice inversion center have  received intensive
attentions for the possibility of spin-triplet dominated pairing
symmetry. An important example is noncentrosymmetric superconductor
CePt$_{3}$Si\cite{CePtSi}. Due to the nontrivial antisymmetric
spin-orbit coupling(ASOC) effect induced by the heavy Platinum atom
and the lacking of inversion symmetry, unconventional
superconducting properties are observed, for instance the high upper
critical field far beyond the Pauli-Clogston
limit\cite{paulilimite}. The subsequent nuclear spin-lattice
relaxation rate as well as magnetic penetration depth measurements
show line nodes in superconducting gap for
CePt$_{3}$Si\cite{CePtSinode1,CePtSinode2}. For a material without
inversion symmetry the spin degeneracy is lifted by ASOC, under such
a condition, orbital angular momentum($\hat{L}$) and spin angular
momentum($\hat{S}$) are not good quantum numbers any more, thus the
strict categorization of even-parity spin singlet and odd-parity
spin triplet conformed to Pauli's exclusion and parity conservation
is violated, then spin triplet may be mixed with spin singlet. Up to
now, several noncentrosymmetric superconductors have been reported,
the development concerning such a scope could be consulted  a
overview given by Sigrist et. al.\cite{noncentrosymmetric}. However
strong electronic correlation in some materials complicates the
studies on ASOC effect, such as heavy fermion properties of
CePt$_{3}$Si. Recently the pairing symmetry of Li$_{2}$Pt$_{3}$B has
been proved to be with line nodes\cite{LiPtBnode}, such a material
is not strongly correlated and could be regarded as an appropriate
example for ASOC effect. For more comprehension on ASOC, thus, any
efforts searching for new noncentrosymmetric superconductors are
worthwhile, especially in the absence of  correlation effect.

In the present paper we report  a  new noncentrosymmetric
superconductor Ru$_7$B$_3$ with $T_{C}$ equal to 3.3 K. Transport
measurements as well as electronic band structure calculation gave a
detailed description to the properties. As to our best knowledge,
neither measurements nor calculations have been embarked except
Matthias mentioned only the $T_{C}$ as 2.58 K in
1950's\cite{Matthias}. Additionally, for the first time we point out
the noncentrosymmetric characterization of Ru$_7$B$_3$.

The paper is organized as follows: Section II describes the sample
preparation process and the  experimental details; then in section
III we provide the structure illustration and the measurement of
many transport properties, including AC magnetization, resistivity,
magnetoresistance and Hall coefficient; section IV focuses on the
specific heat and lower critical field measurements; in the
following section V full potential electronic structure calculation
is discussed; finally a self-consistent normal state and
 superconducting parameters are determined in section VI.

\section{Experiment}
Polycrystalline sample Ru$_7$B$_3$ was synthesized by conventional
solid state sintering process. Stoichiometric Ruthenium
powder(purity 99.9\%) and Boron powder(purity 99.95\%) were mixed
together and grounded thoroughly, then the mixture was pressed into
 pellet with a pressure 10 MPa. The pellet was wrapped with tantalum
foil and sealed into an evacuated silicon tube. 95\% high purity Ar
mixed by 5\% H$_{2}$ was selected as protected atmosphere. The tube
was warmed up to 1000 $^{\circ}$ C in a muffle furnace with a rate
100 $^{\circ}$ C/h and sintered for 100 hours for composition
homogeneity, then the furnace was cooled down to room temperature.
The ingot was ceramic-like and with a silver gray color. It was
observed that the superconducting quality was stable and not
sensitive to humidity or oxygen.  X-ray diffraction (XRD) pattern
measurement was performed at room temperature employing an M18AHF
x-ray diffractometer (MAC Science). Cu $K_{\alpha}$ was used as the
radiation target. Crystallographic orientation and index were
determined by Powder-X\cite{powderX}, an software for processing
x-ray diffraction data. The AC susceptibility were measured based on
an Oxford cryogenic system (Maglab-Exa-12). The resistivity and
specific heat were measured on the Quantum Design instrument PPMS
with temperature down to 1.8 K and the PPMS based dilution
refrigerator (DR) down to 50 mK. The temperatures of both systems
have been well calibrated showing consistency with an error below
2\% in the temperature range from 1.8 K to 10 K. The temperature
dependence of magnetization was measured on the Quantum Design
instrument MPMS with temperatures down to 2 K. The lower critical
field measurement was based on a two dimensional electron gas(2DEG)
micro Hall probe with an active area of 10$\times$10 $\mu$m$^{2}$.
All M(H) curves were taken in zero field cooled mode with initial
temperature up to 6 K. A low field sweep rate of 60 Oe/min was
selected to measure isothermal magnetization curves. Self-consistent
band calculations were carried out using the linear muffintin
orbital(LMTO) method on full potential plane wave
representation(FP-LMTO-PLW)\cite{FP-LMTO-PLW}.

\begin{figure}
\includegraphics[width=8cm]{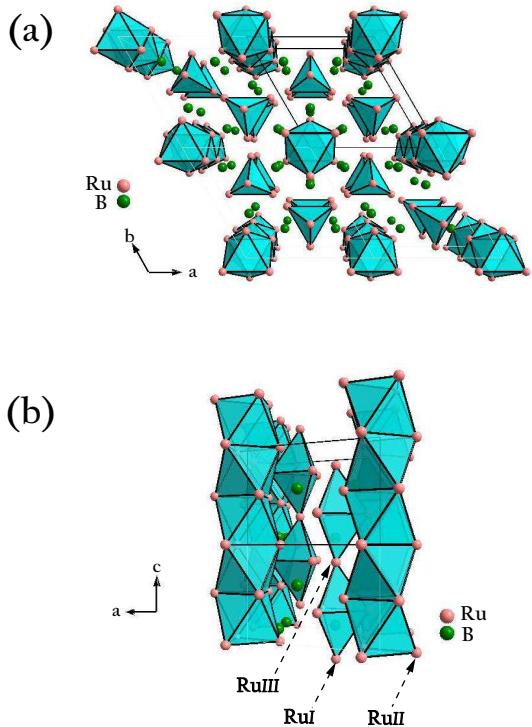}
\caption{(a) a-b projection of Ru$_7$B$_3$ structure, the skeleton
is consisted of  metal tetrahedra and metal octahedra and
interstitial Boron anions. From such a projection, the lattice with
hexagonal rotation symmetry is obvious. (b) a-c projection of the
lattice, two 'chains' are built up by ruthenium ions at different
coordinations. } \label{fig1}
\end{figure}

\section{Structure illustration and transport properties}
The crystal structure of Ru$_7$B$_3$ was determined by
Aronsson\cite{Aronsson} in the late 1950's. It was found that the
lattice is hexagonal with a space group $P6_{3}mc$. There are 20
atoms in one unit cell with effective coordinate Ru$_{I}$(6c),
Ru$_{II}$(6c), Ru$_{III}$(2b) and B(6c), respectively. Thus two
formula units exist in one unit cell. In that paper the author gave
a relative comprehensive description to the lattice structure of
Ru$_7$B$_3$, however, thanks to the complex structure of transition
metal boride, a more detailed illustration should be added as a
supplement for better comprehension on it's properties. As to space
group $P6_{3}mc$, a distinct characterization of the crystal lattice
is without inversion symmetry, for example, the center of  Boron
atoms' sub-lattice dislocates the counterpart of Ruthenium atoms
along $\textit{c}$-axis, thus the inversion symmetry is broken along
such direction. For more clear understanding on the crystal lattice,
we illustrate the structure along two types projection in Fig. 1.
Fig. 1(a) shows the ab projection of Ru$_7$B$_3$ structure. It is
found that the skeleton consists of metal tetrahedra and metal
octahedra and interstitial Boron anions. From such a projection, the
lattice with hexagonal rotation symmetry is obvious. Fig. 1(b) is
the ac projection illustration. A very interesting phenomenon is
that two 'chains' are built up by ruthenium ions at different
coordinations. The metal octahedra at each corner of the lattice is
built up by Ru$_{II}$(6c),  those octahedras share face along c-axis
and thus a zig-zag chain is formed as shown in Fig. 1(b). The rest
Ru$_{I}$(6c) and Ru$_{III}$(2b) form two tetrahedras in one unit
cell at different (x,y) positions, along $\textit{c}$ direction two
tetrahedras (at the same (x,y) positions) share face and then
forming a hexahedra, each hexahedra is connected by Ru$_{III}$(2b)
and thus another type of row is formed. The special structure
configuration might play an important role in transport properties,
also the environment of Ruthenium ions (including the interatomic
distance and the number of the nearest boron ions) is very
important. Detailed parameters are included in Table 1 as shown
below.

\begin{table}
\caption{Structure parameters of  Ru$_7$B$_3$.}
\begin{tabular*}{9cm}{@{\extracolsep{\fill}}cccccc}
\hline \hline
atom & cite    & x   &  z   & No.$^{a}$ & $l$ (${\AA}$)$^{b}$\\
\hline
Ru$_{I}$     &     6c    &   0.4563    &   0.318  & 4 & 2.15, 2.15, 2.66, 2.66\\
Ru$_{II}$    &     6c    &   0.1219    &   0      & 4 & 2.15, 2.16, 2.16, 2.86\\
Ru$_{III}$   &     2b    &   1/3       &   0.818  & 3 & 2.20, 2.20, 2.20 \\
B            &     6c    &   0.187     &   0.582  & / & /  \\

\hline \hline
\end{tabular*}
\label{table} \footnotetext[1]{No. represents the number of nearest
Boron for different cites of Ruthenium.} \label{table}
\footnotetext[2]{$l$ stands for the interatomic distance between
Boron ions and Ruthenium ions.}
\end{table}

Fig. 2  shows the XRD pattern of the sample Ru$_7$B$_3$, which can
be indexed in a hexagonal symmetry with \emph{a} = \emph{b} = 7.4629
${\AA}$ and \emph{c} = 4.7141 ${\AA}$. The indexed indices slightly
deviate from the reported  parameters  \emph{a} = \emph{b} = 7.467
${\AA}$ and \emph{c} = 4.713 ${\AA}$. It is clearly found that most
diffraction peaks are indexed except one minor peak possibly from
un-reacted boron.  From the quality of XRD data, it is estimated
that the purity of Ru$_7$B$_3$ we synthesized is about 95\%. In the
following specific heat measurement, 90\% superconducting component
also prove the sample's purity.

\begin{figure}
\includegraphics[width=8cm]{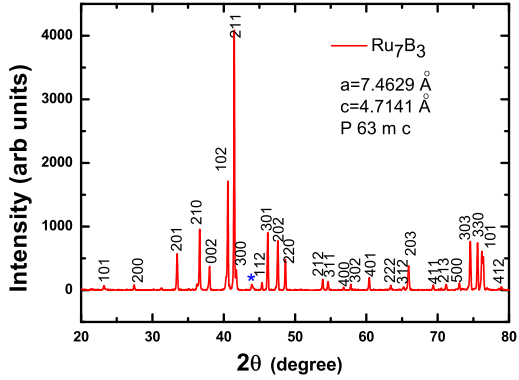}
\caption {(Color online) X-ray diffraction of Ru$_7$B$_3$, most
diffraction peaks are indexed except one minor peak possibly from
un-reacted boron, the purity is estimated about 95\%. The asterisk
marks the peak from impurity phase.} \label{fig2}
\end{figure}

Fig. 3 presents the AC susceptibility measurement under different
magnetic fields. It is found that a sharp diamagnetic transition
happens at 3.3 K and the  diamagnetic value approaches a constant at
3K, 0.3 K transition width indicates the good superconducting
quality. Applying magnetic fields, the transition curve moves
parallel to low temperatures. For estimating the upper critical
field, we take 95\% diamagnetic value as criterion and will discuss
the result in the following paragraph.

Fig. 4 shows the resistivity data from 2 K to room temperatures, the
curve shows a good metallic behavior with a zero temperature
resistivity($\rho_{0}$) 9 $\mu\Omega\cdot$cm. Such a high
conductivity is the common feature of transition metal boride. It is
found that the residual resistivity ratio(RRR) is 28, the relative
high RRR in polycrystalline sample indicates that impurity
scattering is trivial for conductivity. The inset of Fig. 4 shows
the enlargement of superconducting transition, the resistivity drops
sharply  to zero at 3.3 K. Thus the resistivity data and AC
susceptibility give a self-consistent T$_{C}$=3.3 K for Ru$_7$B$_3$,
which is slightly higher than the preliminary mentioned
$T_{C}$$\approx$2.58 K by Matthias\cite{Matthias}. Considering the
high purity of the sample, a quantitative analysis of normal state
resistivity is deserved. We try Wilson's model  for transition
metals\cite{Wilsonmodel}

\begin{equation}
\rho(T)=\rho_{sd}=\kappa_{sd}[\frac{T}{\Theta_{sd}}]^{3}\int_{0}^{\Theta_{sd}/T}\frac{x^{3}dx}{(e^{x}-1)(1-e^{-x})},
\end{equation}
where $\Theta_{sd}$ is a cutoff similar to the Debye temperature and
$\kappa_{sd}$ is a constant. Using  parameters $\Theta_{sd}$=500 K
and $\kappa_{sd}$=1350, a good fitting is obtained as shown in Fig.
4.

\begin{figure}
\includegraphics[width=8cm]{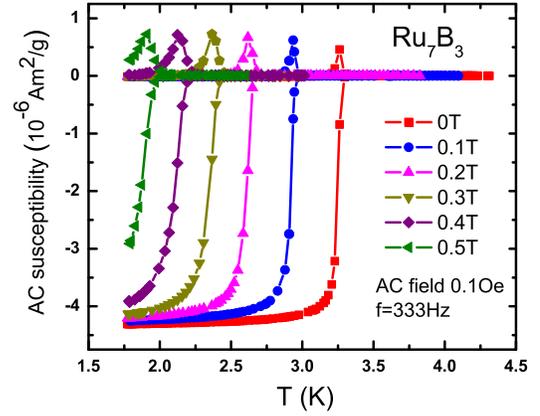}
\caption{(Color online) AC susceptibility of Ru$_7$B$_3$, a sharp
superconducting transition happens at 3.3 K under zero field, a
narrow transition width less than 0.3 K indicates the good
superconducting quality.} \label{fig3}
\end{figure}

\begin{figure}
\includegraphics[width=8cm]{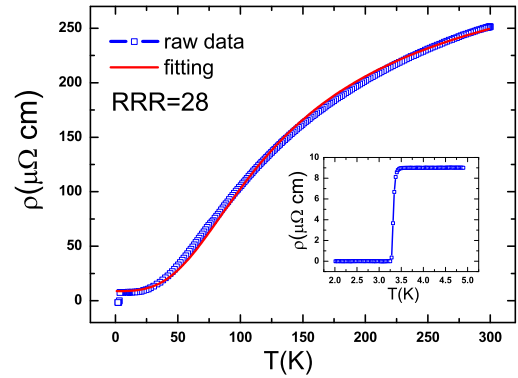}
\caption{(Color online) Resistivity of Ru$_7$B$_3$  under zero field
is measured from 2 K to room temperatures, a good fitting is given
by Wilson s-d scattering model. The inset shows the enlargement of
superconducting transition, the transition width is less than 0.3
K.} \label{fig4}
\end{figure}

Fig. 5 shows the field-broadening resistivity measurement on
polycrystalline Ru$_7$B$_3$ down to 100 mK. When magnetic field is
applied, the onset parts of transition is rounded very similar to
the feature of critical fluctuation in superconductors. While adding
fields up to 0.6 Tesla, the rounded part of transition evolves into
a 'kink' structure as shown clearly in Fig. 5. It seems that the
kinks break the superconducting transition curves into two parts,
the lower part moves  quickly  to low temperatures showing a strong
dependence of magnetic fields, while the  upper parts respond
reversely. It is observed that the zero resistivity point approach
zero temperature when sample is bearing 1.1 T magnetic field, while
superconductivity is depressed completely at about 5 T.  It is thus
very interesting to note that
 the  determined upper critical field strongly depends on the selected resistive criterion.
 As shown in Fig. 6, the criterion of 99\% $\rho_{n}$ and zero resistivity will lead to two distinct  H$_{C2}$(0)
 with a ratio of about 5. A conventional interpretation, as discussed in
Mg$_{10}$Ir$_{19}$B$_{6}$\cite{MgIrB}, for field broadening
transition is filamentary-like superconductivity along grain
boundaries, the stronger scattering reduces the mean free path and
consequently influences on the coherence length. However, we argue
that unlike the case of Mg$_{10}$Ir$_{19}$B$_{6}$ the clear kink
feature in Ru$_7$B$_3$ complicates the determination for H$_{C2}$,
also the kink doesn't come from sample inhomogeneity or two phases
because of the sharp transition in low fields. As to our knowledge,
the unconventional `kink' or `step' features in transition curves
have been extensively discussed in single crystals MgB$_{2}$.
Several reasons have been attributed to this effect, including
superconducting fluctuation\cite{scfluctuation}, two superconducting
gaps\cite{two-gap}, surface barriers\cite{SB} and vortex lattice
melting\cite{vortexmelting1,vortexmelting2},etc. However taking
account of the very low $T_{C}$ and H$_{C2}$, superconducting
fluctuation is believed to be weak. For surface barriers and vortex
lattice melting, the polycrystalline quality seems to exclude their
possibilities. As to the two-gap scenario, the following specific
heat and lower critical field measurements oppose such a point of
view. Thus we assumed the possibility that spin triplet induced kink
feature in the framework of inversion symmetry is broken. It is
known that applying fields broke time reversal symmetry and is
detrimental to spin singlet, whereas triplet pairing remains
unaffected. So for a superconductor with pairing symmetry mixed by
singlet and triplet, the kink feature in field broadening
resistivity curves could be plausible on certain extent. In Fig. 6
we plot the phase diagram of Ru$_7$B$_3$, the criteria are taken as
below, for resistivity 99\%$\rho_{n}$ and zero resistance, for
magnetization 95\% diamagnetic signal and the half position of
specific heat anomaly for thermodynamic measurement. Thus the
derived d$H_{C2}$/d$T$ equal to -0.43 T/K or -0.277 T/K for criteria
99\%$\rho_{n}$ and zero resistance, respectively. It is found that
except 99\%$\rho_{n}$ the other three criteria determined data
points overlap almost together. Further consideration is that bulk
properties provided by specific heat measurement, thus in the
following discussion we use 1.1 T as H$_{C2}$(0) and -0.277 T/K as
d$H_{C2}$/d$T$ near T$_{C}$. We also try to use Ginzburg-Landau
formula to fit the data points determined by 99\%$\rho_{n}$,

\begin{equation}
H_{C2}(T)=H_{C2}(0)\frac{1-t^{2}}{1+t^{2}},
\end{equation}
where t is the normalized temperature T/T$_{C}$, it is found that
the fitting curve strongly deviates the 99\%$\rho_{n}$ points as
shown in Fig. 6, indicating the invalidity of Landau twice order
phase transition theory in the present material.

\begin{figure}
\includegraphics[width=8cm]{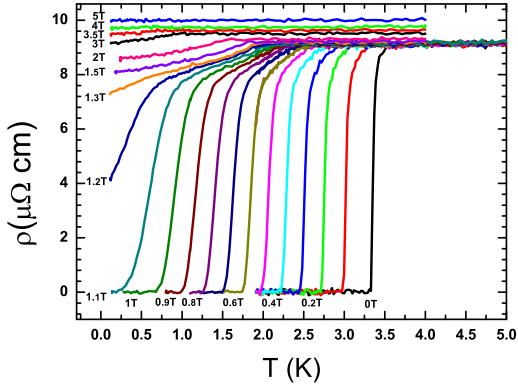}
\caption{(Color online)Field broadening superconducting transition
curves of Ru$_7$B$_3$ from 0.1 K to 4 K, a clear kink feature
appears as the field  exceeds 0.6 T. The drop of resistivity is
totally suppressed by 5 T magnetic field. Distinct
magnetroresistance is observed.  } \label{fig5}
\end{figure}

\begin{figure}
\includegraphics[width=8cm]{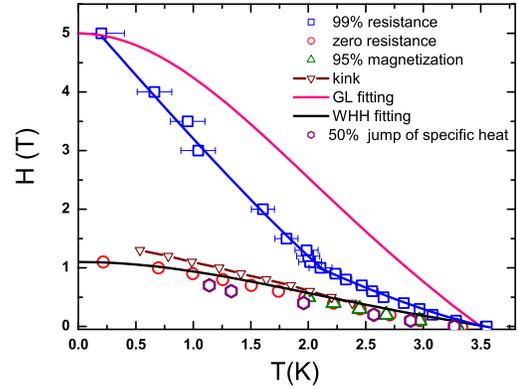}
\caption{(Color online)  Phase diagram of Ru$_7$B$_3$, a non-trivial
flux flow area for criteria 99\%$\rho_{n}$ and zero resistance is
formed by kink shape of transition curves. 95\% diamagnetic signal
from AC susceptibility and one half points of specific heat jumps
are also selected as criteria to estimate the intrinsic value of
$H_{C2}$(0). It is shown that except for  the criterion of
99\%$\rho_{n}$ the other points from different criteria overlap
together, thus the experimental data 1.1 T  is determined as
$H_{C2}$(0). It is found that the estimation of $H_{C2}$ from WHH
formula fits experimental data very well, whereas Ginzburg-Landau
formula fails to describe the upper bound of $H_{C2}$.} \label{fig6}
\end{figure}

Another distinct characterization of Fig. 5 is the field induced
magnetoresistance. Typically magnetoresistance is used to
investigate the electronic scattering process and provide useful
information on  fermi surface (FS) topology. So detailed studies are
needed. In Fig. 7 we present the temperature dependence of
resistivity under magnetic fields from 0 to 9 T. It is found that
the magnetoresistance(MR) is about 16\%
[($\rho_{9T}-\rho_{0T}$)/$\rho_{0T}$] at 5 K, which is one order of
magnitude larger than the ratio of recently discovered iron arsenide
LaFeAsO$_{1-x}$F$_{x}$\cite{MRLaFeAsO}. The latter was regraded as
superconductor with multiple bands. A simple verification  for the
possibility of multigap effect is the scaling based on Kohler's
rule. The Kohler's rule is written as
$\Delta\rho/\rho_{0}$=F(H/$\rho_{0}$), where F is a function
depending on the nature of the metal itself. For single band metal
with symmetric Fermi surface topology Kohler's law should be
conserved. It is shown as Fig. 7(b) that Kohler's rule is only
slightly violated. Unlike typical multi-band superconductor
MgB$_{2}$\cite{MgB2} and LaFeAsO$_{1-x}$F$_{x}$, the breakdown of
Kohler's law is trivial in  Ru$_7$B$_3$, indicating a dominated
single band behavior. The further specific heat and lower critical
field measurements further provide the same conclusion. However we
believe that the slightly violation of Kohler's rule could be
induced by noncentrosymmetric structure of Ru$_7$B$_3$, due to ASOC,
the degenerate spin-up and spin-down bands are spilt, so Kohler's
rule is slightly broken.

\begin{figure}
\includegraphics[width=8cm]{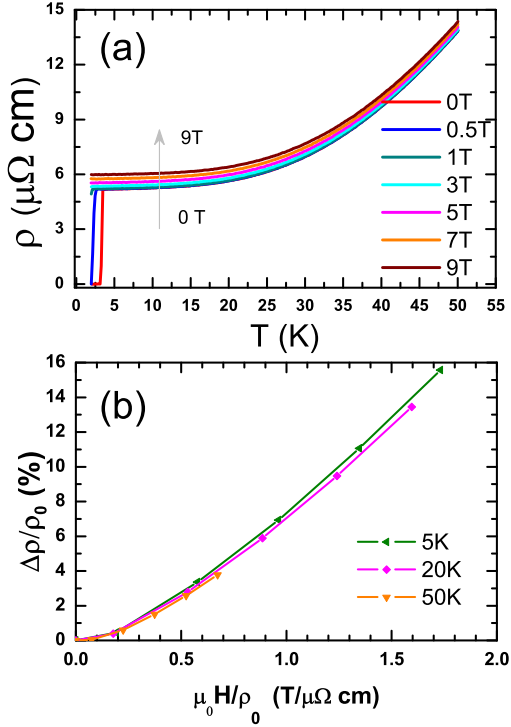}
\caption{(Color online) (a) Temperature dependence of resistivity of
Ru$_7$B$_3$ under magnetic fields from 0 to 9 T, (b) the derived
magnetroresistance $\Delta\rho$/$\rho_{0}$ is 16\% at 5 K, it is
shown that the scaling of Kohler's law is slightly violated. The
violation is attributed to band splitting by ASOC instead of
multiband effect.} \label{fig7}
\end{figure}

Hall coefficient (R$_{H}$) measurement was done by sweeping
temperature at  magnetic field 9 T and reversing field (-9 T). For
avoiding the possible temperature hysteresis, increasing temperature
mode with a moderate rate 1 K/min was adopted for both positive and
negative fields. The Hall coefficient is shown in Fig. 8, it is
found that the charge carrier of Ru$_7$B$_3$ is dominated by
hole-like carriers with R$_{H}$ 3$\thicksim$6$\times10^{-10}$
$m^{3}/C$ from 2 K to 200 K. For verifying the R$_{H}$, we also use
the values derived from sweeping field at three temperature points 2
K, 100 K and 200 K. The low temperature R$_{H}$ is consistent with
the value from sweeping temperature, while error bars exist in high
temperatures. The charge carrier density calculated by
n=1/(R$_{H}$$\cdot$e) is about 1$\times10^{22}/cm^{3}$, which is two
order of magnitude larger than low carrier density superconductors,
for example, cuprates\cite{necuprate} and hole doped iron arsenide
(La$_{1-x}$Sr$_{x}$)FeAsO\cite{neLaSr}. We also notice the nonlinear
 temperature dependence of R$_{H}$, however, the relative change of R$_{H}$ from 2 K to 200 K is small
as shown in Fig. 8. Moreover, it is known that the hall effect is
very sensitive to the temperature dependent scattering rate, local
fermi velocity and complex FS topology\cite{Hallsensitive}, thus
considering the polycrystalline quality detailed analysis on
temperature dependent Hall coefficient would not be discussed here.

\begin{figure}
\includegraphics[width=8cm]{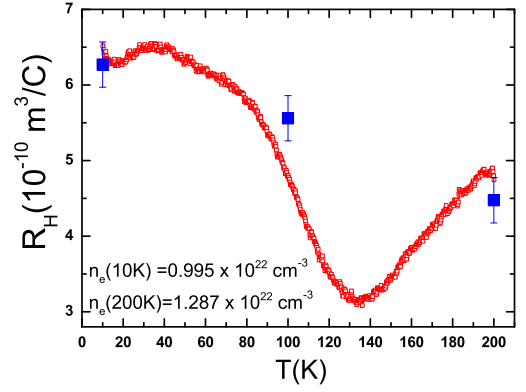}
\caption{(Color online)Hall coefficient measurement of Ru$_7$B$_3$
from sweeping temperature under reversing fields, the charge carrier
is dominated by hole from low temperature to 200 K, the density of
charge carrier is about $1\times10^{22}/cm^{-3}$. The squares
represent the data measured by sweeping fields at fixed
temperatures. } \label{fig8}
\end{figure}

\section{Specific heat and lower critical field measurement}

\subsection{Specific heat}
 Fig. 9 shows the raw data of specific heat
under different magnetic fields from zero to 3 T. With increasing
fields the specific heat anomaly near T$_{C}$ move quickly to low
temperatures leaving a background consistent with  that above
T$_{c}$ at zero field. Thus the normal state specific heat could be
extracted easily with the relation $C/T=\gamma_n +\beta T^2$, where
$\gamma_n$ is the normal state specific heat coefficient and $\beta$
corresponds to phonon contribution. It is found that $\beta$=0.3732
$mJ/ mol K^4$ and $\gamma_n$= 89.95 $mJ/mol K^2$, however a residual
value $\gamma_0 \approx$ 9.8  $mJ/ mol K^2$ indicates the existence
of about 10\% non-superconducting fraction. The non-superconducting
fraction could partly come from unreacted boron as inferred from
analysis from X-ray diffraction. Thus the normal state Sommerfeld
constant could be determined from the relation $\gamma
_{e}$=$\gamma_n$-$\gamma_0$ as 80.15 $mJ/mol K^2$. Using the
relation $\Theta_D$ =$(12\pi^4k_BN_AZ/5\beta)^{1/3}$, where $N_A$ =
6.02$\times 10^{23}$ the Avogadro constant, Z=20 the number of atoms
in one unit cell, we get the Debye temperature $\Theta_D $ = 470.18
K. In the previous section, we obtained a similar value
$\Theta_{sd}$=500 K from the resistivity curve fitting, the
consistent values prove the reliability of two different
measurements. It is noticed that comparing with the values of
Mg$_{10}$Ir$_{19}$B$_{6}$\cite{MgIrBwen} and
Li$_{2}$Pt$_{3}$B\cite{LiPtBSH} $\gamma_n$ in our measurement is
relatively high, thus a prudent checking is necessary. For a type-II
superconductor, $\gamma_n$ could be estimated as the following
relation\cite{gamma-n}:

\begin{equation}
-\frac{\partial \mu_0H_{C2}}{\partial T}|_{T_c}=A \rho_n\gamma_n
\eta,\label{eq:2}
\end{equation}
where $A=3.81e/\pi^2k_B$=4479.21$(T/K)(\Omega m)^{-1}(J/m^3
K^2)^{-1}$, using $- \partial \mu_0H_{C2}(T)/\partial T|_{T_c}
\approx$ 0.277 T/K, $\rho_n$ = 9 $\mu\Omega cm$, and taking $\eta
=1$ for the weak coupling case, we have $\gamma_n$ = 94 $mJ/mol K^2$
which is very  close to the upper bound of the experimental value
89.95 $mJ/mol K^2$. Further we could  estimate the $\lambda_{e-Ph}$
 via McMillan equation\cite{McMillan}:

\begin{equation}
T_{C}=\frac{\Theta_{D}}{1.45}exp(-\frac{1.04(1+\lambda_{e-Ph})}{\lambda_{e-Ph}-\mu^{\ast}(1+0.62\lambda_{e-Ph})}),
\end{equation}
where $\mu^{\ast}$ is the Coulomb pseudopotential taking 0.11,
$\Theta_{D}$=470.175 K and T$_{C}$=3.3 K, we obtain
$\lambda_{e-Ph}$=0.48. The value indicates that Ru$_{7}$B$_{3}$
belongs to a weak coupling superconductors.

\begin{figure}
\includegraphics[width=9cm]{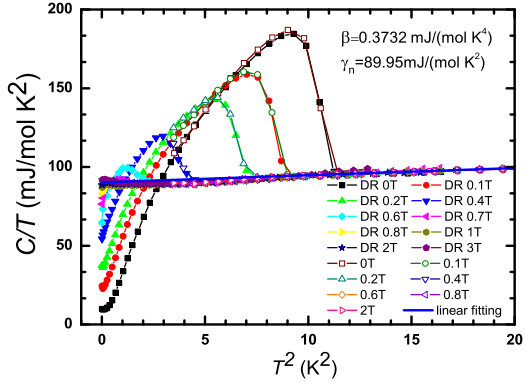}
\caption{(Color online) Raw data of specific heat plotted as $C/T$
vs. $T^2$. All filled symbols represent the data taken with the DR
based on the PPMS at various magnetic fields. The open squares show
the data taken with the PPMS at different fields. The thick solid
line represents the normal state specific heat which contains both
the phonon $\gamma_{ph}$ and the electronic contributions.}
\label{fig9}
\end{figure}

For noncentrosymmetric superconductors, novel pairing symmetry could
be achieved due to the mixing of spin singlet and spin triplet.
Specific heat is a useful tool to investigate the material's low
energy excitation. So, subtracting the contribution of phonon, we
present temperature dependence of $\gamma_{e}$ under magnetic fields
up to 3 T in Fig. 10(a). For the convenience of theoretical
analysis, we further subtract $\gamma_n$ of zero field data as shown
in Fig. 10(b). Thus  the weak coupling BCS formula could be used:

\begin{eqnarray}
\gamma_e(T) = \frac{4N(0)}{k_{B}T^{3}}
\int^{\hbar\varpi_{D}}_{0}\int^{2\pi}_{0}\frac{e^{\xi/k_{B}T}}{(1+e^{\xi/k_{B}T})^{2}}\nonumber\\
\times(\varepsilon^{2}+\Delta^{2}(\theta,T)-\frac{T}{2}\frac{d\Delta^{2}(\theta,T)}{dT})d\theta
d\varepsilon,\label{eq:5}
\end{eqnarray}
where $\zeta=\sqrt{\varepsilon^2+\Delta^2(T,\theta)}$. In obtaining
the theoretical fit we take the implicit relation $\Delta_{0}(T)$
derived from the weak coupling BCS theory for superconductors with
different pairing symmetries: $\Delta(T,\theta)=\Delta_{0}(T)$ for
s-wave, $\Delta(T,\theta)=\Delta_{0}(T)\cos 2\theta$ for d-wave, and
$\Delta(T,\theta)=\Delta_{0}(T)\cos \theta$ for p-wave,
respectively. The theoretical curve of s wave fits the experimental
data very well leading to an isotropic gap value $\Delta_0$ = 0.5
$meV$ and $T_c$ = 3.22  $K$. The ratio $\Delta_0/k_BT_c$ = 1.80
obtained here is quite close to the prediction for the weak coupling
limit ($\Delta_0/k_BT_c$ = 1.76). In addition, the specific heat
anomaly at T$_{c}$ is $\Delta C_e/\gamma_nT|_{T_c} \approx $ 1.31
being very close to the theoretical value 1.43 predicted for the
case of weak coupling.

\begin{figure}
\includegraphics[width=7cm]{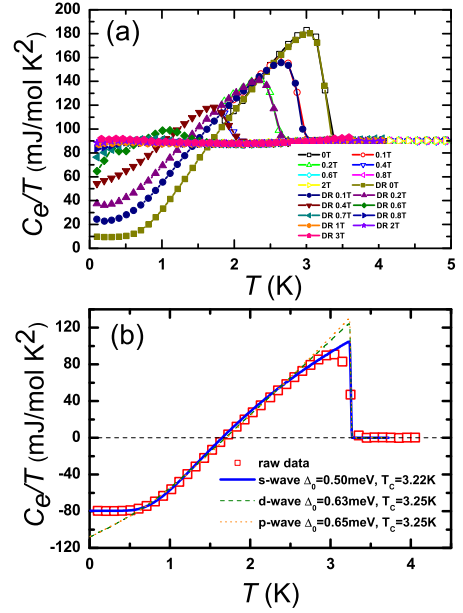}
\caption{(Color online) (a) Temperature dependence of $\gamma_{e}$
for magnetic field up to 3 T, (b) Temperature dependence of
$\gamma_{ne}$-$\gamma_{e}$ at zero field. The blue solid, green
dashed and orange dotted lines are theoretical curves calculated
based on BCS model with a gap of s wave, d wave and p wave,
respectively.} \label{fig10}
\end{figure}

Condensation energy($E_{C}$) is an important parameter for
superconductor, thus we  calculate $E_{C}$ with the following
process, firstly the entropy difference between normal state and
superconducting state could be obtained by
$S_n-S_s=\int_0^T(\gamma_n-\gamma_e)d T'$, and then E$_{C}$ is
calculated through $E_{c}=\int_T^{4K}(S_n-S_s)dT'$. The resulted
temperature dependence of  E$_{C}$ is shown in Fig. 11, the inset is
the entropy difference between normal state and superconducting
state. The $E_{C}$ is about 192 mJ/mol at 0 K. Alternatively,
$E_{C}$ could be calculated by the following equation:

\begin{equation}
E_{c}=\alpha N(E_F)\Delta_0^2/2=\alpha
\frac{3}{4\pi^2}\frac{1}{k_B^2}\gamma_n\Delta_0^2,\label{eq:4}
\end{equation}

For a BCS s-wave superconductor, $\alpha$$\approx$1, taking
$\gamma_n$=80.15 mJ/mol$\cdot$K$^{2}$ and $\Delta_{0}$=0.5 meV, we
obtain a value 205 mJ/mol, which is close to experimental value 192
mJ/mol. In addition, the consistence reversely verifies the validity
of $\gamma_n$ and $\Delta_{0}$ determined through our experiment.
From condensation energy, the thermodynamic critical field
$\mu_{0}$H$_{C}$(0) could be calculated via the relation
$\mu_{0}H_{C}^{2}(0)/2=F_{N}-F_{S}=\iint (\gamma_{n}-\gamma_{e})dT$,
yielding $\mu_{0}$H$_{C}$(0)= 612 Oe. For a comparison,
$\mu_{0}$H$_{C}$(0) of another noncentrosymmetric superconductor
Mg$_{10}$Ir$_{19}$B$_{6}$ is about 300 Oe\cite{MgIrB}.

\begin{figure}
\includegraphics[width=8cm]{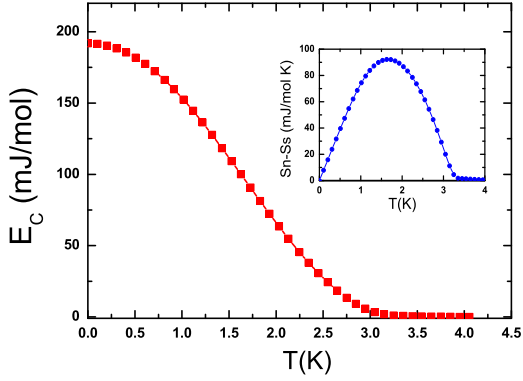}
\caption{(Color online) Superconducting condensation energy of
Ru$_7$B$_3$ calculated by specific heat, the inset shows entropy
difference between the normal and superconducting state. }
\label{fig11}
\end{figure}

\subsection{Lower critical field measurement}
Lower critical field(H$_{C1}$) is an important parameter for a
superconductor. According to the Ginzburg-Landau theory, H$_{C1}$
reflects the superfluid density $\rho_{s}$ since H$_{C1}$ is related
to London penetration depth $\lambda$ and thus a relation is
established that H$_{C1}$$\sim$1/$\lambda$$^{2}$. Moreover, the
temperature dependence of H$_{C1}$, especially the low temperature
features, is always used to investigate the superconducting pairing
symmetry and multi-gap effect, for instance the node feature for
pairing symmetry and gaps' value for the latter. In this section we
used a two dimensional electron gas(2DEG) micro Hall probe to
measure $\mathbf{local}$ magnetization loops of Ru$_7$B$_3$. For
weakening the complex effects of the character of field penetration,
such as Bean-Livingston surface barriers and geometrical barriers,
we used a low field sweep rate of 60 Oe/min to measure isothermal
magnetization curves.

\begin{figure}
\includegraphics[width=8cm]{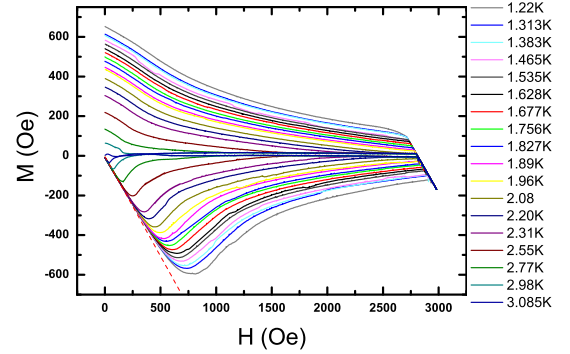}
\caption{(Color online) The raw data of M(H) curves at different
temperature, The red dashed line give the Meissner linear approach.
It is found that the value of M is very approaching to that of
magnetic field, indicating almost 100\% diamagnetization at lower
temperatures. } \label{fig12}
\end{figure}

Fig. 12 is the initial isothermal M(H) curves over the temperature
range from 1.22 K to 3.2 K. It is found that the low-field parts of
those M(H) curves overlap almost on one line (red dash line is
guided to the eyes in Fig. 12) with a constant slope, which is
attributed to Meissner effect and called as Meissner line. Thus
H$_{C1}$ could be determined as the departure between M(H) curve and
Meissner line with the same criterion for all curves. The resulted
temperature dependence of H$_{C1}$(T)(normalized to H$_{C1}$(0))  is
shown in Fig. 13. The inset of Fig. 13 shows the criterion for
determination of the value of H$_{C1}$ at 1.4 K, a value of 10
 Oe was regarded as error bar due to noise induced uncertainty.
According to BCS theory for clean superconductors, the normalized
H$_{C1}$(T)/H$_{C1}$(0) is expressed as follows\cite{Tinkham}:

\begin{equation}
\frac{H_{C1}(T)}{H_{C1}(0)}\propto\frac{\lambda^{2}(0)}{\lambda^{2}(T)}=1-2\int^{\infty}_{\Delta(T)}(-\frac{\partial
f(E)}{\partial(E)})D(E)d(E),
\end{equation}

Where $\Delta(T)$ is the BCS superconducting energy gap,
$f(E)$=1/[exp(-E/$k_{B}$T)+1] is the Fermi distribution function,
and D(E)=E/[E$^{2}$-$\Delta$$^{2}$(T)]$^{1/2}$ is the quasiparticle
density of states. We use above equation to fit the experimental
data, H$_{C1}$(0) and $\Delta(0)$ are fitting parameters. It is
found that single gap s-wave pairing could give an appropriate fit
with fitting values H$_{C1}$(0)=110 Oe and $\Delta(0)$=0.5 meV, the
latter is consistent with that of specific heat measurement. Thus
the good consistence indicates the reliability of results determined
by H$_{C1}$ measurement, although the lower temperature (less than
1.2 K) experimental data is absent. A possible argument is that the
nominal H$_{C1}$ obtained from experimental data fitting may not
reflect the true value due to the inevitable surface barrier and
geometrical barrier induced by the polycrystalline quality. Thus in
the following paragraph we will do self-consistent checking from
superconducting parameters determined by other measurements.

\begin{figure}
\includegraphics[width=8cm]{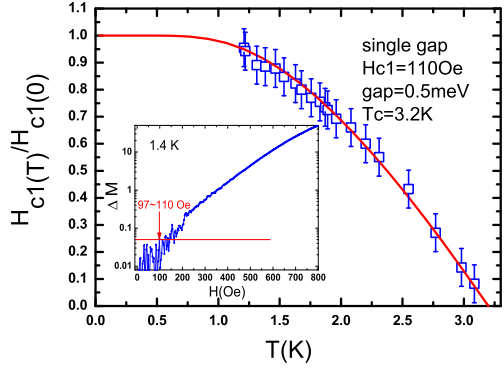}
\caption{(Color online) The extracted $H_{C1}$ as a function of
temperature, isotropic s wave with gap value 0.5 meV could give a
good fitting to the experimental data. The inset shows the criterion
for $H_{C1}$ determination at temperature 1.4 K.} \label{fig13}
\end{figure}

\section{Electronic structure calculation}
In this section we present density of states(DOS) and band
dispersion results based on full potential linear-muffin-tin-orbital
program LMTART by Savrasov\cite{FP-LMTO-PLW,Savrasov}. Full
potential approximation, Plane wave expansion(PLW), is selected and
believed to give the adequate accuracy. Fig. 14 is the DOS
calculation of Ru$_7$B$_3$. The total DOS curve have numerous van
Hove singularities, the feature is very similar to that of
Mg$_{10}$Ir$_{19}$B$_{16}$\cite{MgIrB-DOS}, in that paper the author
attributed the characterization as large numbers of atom in the unit
cell and various interatomic distances. In Ru$_7$B$_3$ only 20 atoms
exist in one unit cell, thus the numerous van Hove singularities
could  stem from the various interatomic distances complicated by
the lacking of a inversion center. Another feature is that
electronic structure is dominated by Ruthenium 4d states, boron 2p
state contributes weakly. It is reasonable to understand from
structure aspect that the lattice of Ru$_7$B$_3$ is mainly consists
of metal tetrahedra and metal octahedra or 'chains' along
$\textit{c}$-direction, thus charge carries naturally favor those
special channels in a crystal lattice. The total DOS at chemical
potential for Ru$_7$B$_3$ is 20.988 state/eV per formula unit. For
checking the calculated DOS at chemical potential, we could simply
estimate the DOS from $\gamma_{n}$ in the framework of free electron
gas.

\begin{equation}
N(0)=(\frac{2\pi^{2} k_{B}^{2}}{3})^{-1}\cdot\gamma_{n},
\end{equation}
where $\gamma_{n}$ is selected as 80.15 $mJ/mol K^2$, k$_{B}$ is
Boltzmann constant=1.380658$\times10^{-23}$ J/K, N(0) represents the
density of state. The obtained N(0) is about 17 state/cell per
formula, which is close the calculated value 20.988. Fig. 15 is the
band dispersion curves near Fermi energy. A distinct feature is that
the all bands are doubly accompanied, the feature is due to
asymmetric spin-orbit coupling effect, thus the degeneracy of
spin-up and spin-down is lifted. It is noticed that at some k-points
with high symmetry splitting instead of degeneracy also exists,
which could come from problems such as un-adequate optimized
parameters at initialization during computation. Nevertheless, it is
believed that such stigma can not affect the main results of the
present paper.

\begin{figure}
\includegraphics[width=8cm]{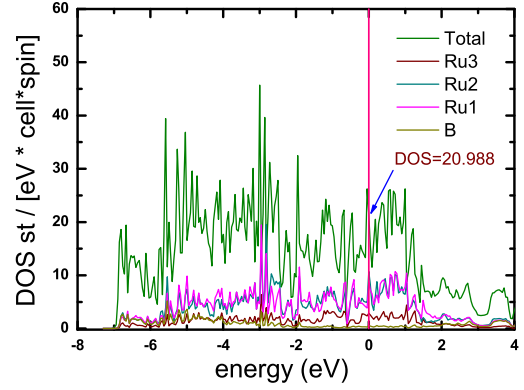}
\caption{(Color online) Density of states of Ru$_7$B$_3$ calculated
by full potential PLW-LMTO, the Dos at chemical potential is about
20.988 state/(eV cell spin). It is shown that electronic structure
is dominated by Ruthenium 4d states, boron 2p state contributes
weakly.} \label{fig14}
\end{figure}

\begin{figure}
\includegraphics[width=8cm]{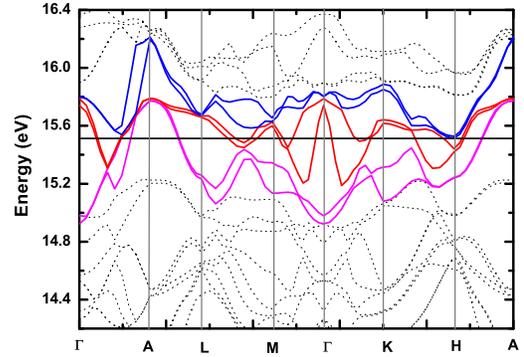}
\caption{(Color online) Band dispersion curves of noncentrosymmetric
materials Ru$_7$B$_3$, the bands are doubly bound due to the ASOC
induced spin-up and spin-down splitting. } \label{fig15}
\end{figure}

\section{Self-consistence among superconducting parameters}
 In the discussion
of magnetoresistance, the Kohler's rule was found to be only
slightly broken, indicating a symmetric Fermi surface topology. Thus
we could deduce the Fermi-wave number($k_{F}$) from charge carrier
density(n)\cite{kF-from-ne} assuming a single spherical Fermi
surface, $k_{F}$=(3$\pi^{2}n)^{1/3}$=6.6527 nm$^{-1}$, where
n=0.995$\times10^{22}$ cm$^{-3}$. The effective mass is estimated as
$m^{\ast}=(3\hbar^{2}\gamma_{n})/(V_{mol}k_{B}^{2}k_{F})$=
17m$_{el}$, where m$_{el}$ is bare electron mass and molar volume
$V_{mol}$=136.89 cm$^{3}$/mol. Then the Fermi velocity
$\nu_{F}$=$\hbar$ $k_{F}$/$m^{\ast}$ is about 0.47$\times$10$^{5}$
m/s. The mean-free-path is evaluated as $l$=$\hbar k_{F}/(\rho_{0} n
e^{2})$=31.36 nm. The superconducting penetration depth
$\lambda(0)=\sqrt{m^{\ast}/\mu_{0}ne^{2}}$ is 214 nm. So the
coherence length could be estimated using the BCS expression
$\xi(0)=0.18\hbar v_{F}/(k_{B} T_{C})$ as 19.5 nm\cite{Tinkham}.
Thus the above superconducting parameters could give a stringent
checking on experimental data, such as $\xi(0)$ and H$_{C1}$ and
$\mu_{0}H_{C}(0)$. In the upper critical fields measurement,
H$_{C2}$(0) is determined as 1.1 T, so using
$\xi(0)$=$\sqrt{\phi_{0}/(2 \pi H_{C2}(0))}$, where $\phi_{0}$ is
flux quanta, coherence length is 17.3 nm, such a value is very close
to the deduced $\xi(0)$ 19.5 nm. For checking on H$_{C1}$,
 the following formula is used\cite{MgIrB}:

\begin{equation}
\mu_{0}H_{C1}=(\frac{\phi_{0}}{4\pi
\lambda_{0}^{2}})ln(\frac{\lambda_{0}}{\xi_{0}}),
\end{equation}
yielding  $\mu_{0}H_{C1}$= 90.2 Oe, where $\lambda_{0}$ and
$\xi_{0}$ are deduced values from charge carrier density. In our
lower critical field measurement we obtain $\mu_{0}$H$_{C1}$=110 Oe,
which is larger than the estimated value 90.222 Oe. The
thermodynamic critical field $\mu_{0}$H$_{C}(0)$ could  be obtained
from the following formula\cite{MgIrB}:

\begin{equation}
H_{C1}H_{C2}=H_{C}^{2}(0)ln(\frac{\lambda_{0}}{\xi_{0}}),
\end{equation}
using H$_{C2}$=11000 Oe, H$_{C1}$=90.2 Oe, $\lambda_{0}$=214 nm and
$\xi_{0}$=17.3 nm, $\mu_{0}$H$_{C}(0)$ is given as 628 Oe, which is
very close to the value of 612 Oe determined by specific heat. For
further checking on the experimental value of H$_{C1}(0)$, firstly
the experimental values $H_{C1}$(110 0e) and $\xi(0)$(17.306 nm) are
used into Equation 9, the solved $\lambda_{0}$=189.3 nm, then using
that value H$_{C}(0)$ could be estimated by Equation 10, yielding
$\mu_{0}$H$_{C}(0)$=711 Oe, which is about 100 Oe larger than that
obtained from specific heat measurement. Therefore it is safe to
conclude that the intrinsic value of H$_{C1}(0)$ is about 90 Oe. So
in Table II we list the superconducting and normal state properties
of noncentrosymmetric material Ru$_7$B$_3$.

\begin{table}
\caption{Superconducting and normal state properties of Ru$_7$B$_3$}
\begin{tabular*}{7cm}{@{\extracolsep{\fill}}cc}
\hline \hline
parameters                                   &  Ru$_{7}$B$_{3}$ \\
\hline
$\gamma_{n}$(mJ/mol K$^{2}$)                 &   80.15   \\
$N(0)$(state/eV cell spin)                     &  20.988 \\
$\Delta$ (m eV)                              & 0.5   \\
$H_{C2}$  (Oe)                               &   11000 \\
$H_{C1}$   (Oe)                              &    90\\
$\xi$  (nm)                                  & 17.3\\
$\lambda$ (nm)                               & 215\\
$\kappa$                                     & 12.4\\
$H_{C}$(0)  (Oe)                             &  612$\sim$628\\
$\Delta$/k$_{B}$T$_{C}$                      &  1.80\\
$\Delta_{c}$/$\gamma_{n}$T$_{C}$             &  1.31\\
$\beta$ (mJ/mol K$^{4}$)                     &  0.3735\\
$\Theta$$_{D}$  (K)                          &  470.175\\
$E_{c}$  (mJ/mol)                            & 192$\sim $205\\
$m^{\ast}$                                  & 17m$_{e}$\\
$n$  (cm$^{-3}$)                         & 1$\times$10$^{22}$\\
$l$ (nm)                                     & 31.36 \\
$\rho_{n}$  ($\mu$$\Omega$$\cdot$ cm)(4 K)        & 9\\

\hline \hline
\end{tabular*}
\label{table}
\end{table}

In the final part of self-consistent checking on physical parameters
of Ru$_7$B$_3$, a brief discussion on the possible exotic properties
due to ASOC is necessary. Generally, two criteria for novel
superconductivity in the framework of noncentrosymmetry have been
established. The first one is Pauli-Clogston limiting
field\cite{paulilimite}, which could be expressed as
$H_{p}(0)=\Delta(0)/2\sqrt{2}\mu_{B}=1.83 T_{C}$, for Ru$_7$B$_3$
with $T_{C}$ 3.3 K, yielding $H_{p}(0)\approx 6$ T. In our
experiment the upper bound of H$_{C2}$(99\%$\rho_{n}$) is found to
be about 5 T, the experimental value is obvious below Pauli limit,
indicating that H$_{C2}$ is still determined by orbital depairing
fields. Nevertheless, the observed kink feature is still a puzzle
under the common thinking, such as two bands or superconducting
fluctuation scenarios, et. al. Another criterion is the presence of
line or point node in the superconducting gap. The good fitting of
specific heat data with isotropic s-wave has provided a strong
evidence that spin singlet dominated the condensate. The lower
critical field measurement also give the same conclusion even though
the lower temperature data (less than 1.2 K) is unfortunately
absent. Thus a safe conclusion could be given that
noncentrosymmetric superconductor Ru$_7$B$_3$ is dominated by s-wave
pairing symmetry, minor spin triplet could admix among the majority
of spin singlet.

\begin{figure}
\includegraphics[width=8cm]{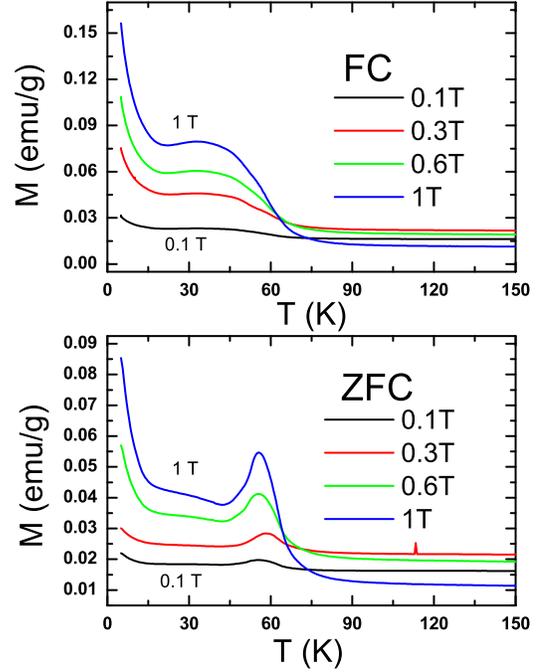}
\caption{(Color online) Normal state magnetization measurement of
Ru$_7$B$_3$ from 5 K to 150 K. (a) FC(field cooling) mode, a rounded
hump happens at about 40 K, (b) ZFC(zero field cooling) mode, the
rounded hump on FC curve changes into peak and moves to higher
temperature about 55 K.} \label{fig16}
\end{figure}

In the end of experiment and discussion, we show magnetization of
Ru$_7$B$_3$ from 5 K to 150 K bearing the magnetic field modes of
FC(field cooling) and ZFC(zero field cooling) in Fig. 16 . It is
found that there is a hump at 40 K on the FC curves, while for ZFC
mode the hump changes into a peak and moves to higher temperature at
about 55 K. For now it is still hard for us to comprehend the normal
state magnetization of Ru$_7$B$_3$, especially  more subtle features
adding on ZFC curves as shown in Fig. 16(b). For comparison we
measured room temperature magnetization of Ruthenium element(not
shown here), it is found that Ruthenium element is paramagnetic at
300 K, lowering temperature an antiferromagnetic transition happens
at 150 K on the paramagnetic background. The density states
calculation have shown that ruthenium contributes most DOS on FS.
Thus from such a point view, the normal state magnetization of
Ru$_7$B$_3$ could be similar to that of ruthenium element.
Furthermore for the ruthenium element the outer shell electron
configuration is $4d^{7}5s^{1}$, thus in Ru$_7$B$_3$ the ruthenium
cations with higher spin angular momentum could be anticipated.
However, we could not exclude the possibility of impurity induced
magnetism due to the 10\% residual $\gamma_{0}$ for specific heat
measurement. Nevertheless, the studies on magnetism of Ru$_7$B$_3$
is worthwhile, for example the possible antiferromagnetic
fluctuation induced superconductivity has been a hot issue in
MgCNi$_{3}$\cite{cava}, for RuSr$_{2}$GdCu$_{2}$O$_{8}$\cite{Tallon}
the interplay between ferromagnetism and superconductor is also very
attractive.

\section{Summary}
Transition metal boride  Ru$_7$B$_3$ was found to be a
noncentrosymmetric superconductor with $T_{C}$ equal to 3.3 K.
Superconducting and normal state properties of Ru$_7$B$_3$ were
determined by a self-consistent analysis through the results of
resistivity($\rho_{xx}$ and $\rho_{xy}$) , specific heat, lower
critical field measurement and electronic structure calculation. It
is found that Ru$_7$B$_3$ belongs to an s-wave dominated single band
superconductor with energy gap 0.5 meV and could be categorized into
type II superconductor with weak electron-phonon coupling. Unusual
'kink' features are clearly observed in field-broadening resistivity
curves, possibly indicating the  admixture of spin singlet and spin
triplet due to the absence of lattice inversion symmetry.

\section{Acknowledgments}

This work is supported by the National Science Foundation of China,
the Ministry of Science and Technology of China (973 project:
2006CB601000 and 2006CB921802), and the Knowledge Innovation Project
of the Chinese Academy of Sciences (ITSNEM). The author thanks to T.
Xiang for helpful discussion and Dr. L. Tang for technical support
on electronic structure calculations. Appreciation also give to C.
Dong for the help of structure analysis.

\end{document}